\documentclass{ptapap}

\author{B.\,Buysschaert}[bram.buysschaert@obspm.fr, LESIA,IvS]
\author{H.\,Pablo}[MONT]
\author{C.\,Neiner}[LESIA]
\affil[LESIA]{LESIA, Observatoire de Paris, PSL Research University, CNRS, Sorbonne Universit\'es, UPMC Univ. Paris 06, Univ. Paris Diderot, Sorbonne Paris Cit\'e, 5 place Jules Janssen, F-92195 Meudon, France}
\affil[IvS]{Instituut voor Sterrenkunde, KU Leuven, Celestijnenlaan 200D, 3001 Leuven, Belgium}
\affil[MONT]{D\'epartement de physique and Centre de Recherche en Astrophysique du Qu\'ebec (CRAQ), Universit\'e de Montr\'eal, C.P. 6128,
Succ. Centre-Ville, Montr\'eal, Qu\'ebec, H3C 3J7, Canada}

\title{Preparing and correcting extracted BRITE observations}

\begin{document}

\maketitle

\begin{abstract}
Extracted BRITE lightcurves must be carefully prepared and corrected for instrumental effects before a scientific analysis can be performed.  Therefore, we have created a suite of Python routines to prepare  and correct the lightcurves, which is publicly available.  In this paper we describe the method and successive steps performed by these routines.

\end{abstract}

\section{Introduction}
The data reduction of the extracted BRITE photometry must take instrumental effects into account \citep{2016arXiv160800282P}.  Therefore, we compiled a suite of Python tools for the preparation and correction of BRITE lightcurves.  These tools are based on our own experience with space-based photometry and the BRITE cookbook\footnote{http://brite.craq-astro.ca/doku.php?id=cookbook} \citep[see also][]{2016A+A...588A..55P}.  These Python routines are publicly available\footnote{http://github.com/bbuysschaert/BRITE\_decor} and can be subdivided into two parts: the preparation of the extracted BRITE photometry, and its decorrelation.

\section{Preparation of BRITE photometry}
Since most of the BRITE photometry will be used for a frequency analysis, having an exact timing of the observations is of paramount importance.  Thus, we first convert the original data timing to a mid-exposure timing.  This permits an easier combination of different BRITE datasets, especially when several exposures are stacked on board or when various exposure times are employed.  The calculated time shift ranges from the sub-second level to tens of seconds and is applied in the satellite reference frame.

Once the timing for all data is adjusted, we clean the data to increase its signal-to-noise ratio by means of outlier rejection.  Not only outlying flux observations are discarded, but we also filtered on the meta-data.  Long-term trends between the (meta-)data and time were accounted for by a LOWESS (locally weighted scatterplot smoothing) filter, prior to the $s\sigma$-clipping or percentage filtering.  Typically, we perform the outlier rejection in the following order: i) Rejection based on the CCD centroid position, simultaneously on both the x and y position;  ii) Discarding data with on-board temperature behaviour not following the long-term trends.  This can occur after an interruption in the observations and affect full orbits;  iii) Flux outliers, which depends heavily on the type of object studied.  iv) Too short satellite orbits, to avoid small number statistics.

\section{Correction of BRITE photometry}
For the correction of the BRITE photometry, we use B-splines to describe the observed correlation.  The knotpoint spacing of the spline as well as its order are treated as free parameters.  The best fit is then quantified by means of a log-likelihood and information criteria \citep{1986ssds.proc..105D, 1990ApJ...364..699A}, which accounts for overfitting.

First, we account for the varying PSF shape due to the fluctuating on-board temperature, possibly enhanced by intrapixel variations.  This instrumental effect leads to a time-dependent correlation between the flux and the PSF shape, and acts mainly as a noise source.  We reduce the dimensionality of this correlation by subdividing the data into shorter time windows, using the long-term temperature behaviour.  Within each time window, the flux is then decorrelated with the CCD centroid positions.  The information criteria determined the sequence of the fits.

Lastly, we perform a traditional decorrelation between the flux and all available meta-data to suppress instrumental signal.  This decorrelation process is applied per BRITE dataset, using the CCD positions and temperature, and the satellite orbital phase.  The normalized correlation matrix indicates the order of the decorrelation process and is recalculated after each iteration.  Each meta-data parameter is only used once for decorrelation and only if the correlation is greater than 10\,\%.

\section{Conclusions}
We developed a suite of routines to prepare and correct extracted BRITE photometry, which is publicly available.  By applying these to the BRITE data of $\zeta$\,Ori (Buysschaert et al. 2016, these proceedings, Buysschaert et al. in prep.), we were able to reduce the noise between a factor 1.2 and 3.5, while also reducing the instrumental signal in the low frequency regime (< 0.05\,c/d) and around 1\,c/d.  Up to now, these routines were only tested on classical BRITE data, but will soon be upgraded for chopping BRITE data.

\bibliographystyle{ptapap}
\bibliography{Buysschaert_BRITEcor}

\begin{thebibliography}{4}
\providecommand{\natexlab}[1]{#1}
\providecommand{\url}[1]{\texttt{#1}}
\providecommand{\urlprefix}{URL }
\providecommand{\eprint}[2][]{\url{#2}}

\bibitem[{{Anderson} et~al.(1990){Anderson}, {Duvall}, \&
  {Jefferies}}]{1990ApJ...364..699A}
{Anderson}, E.~R., {Duvall}, T.~L., Jr., {Jefferies}, S.~M., \emph{{Modeling of
  solar oscillation power spectra}}, \emph{ApJ} \textbf{364}, 699 (1990)

\bibitem[{{Duvall} \& {Harvey}(1986)}]{1986ssds.proc..105D}
{Duvall}, T.~L., Jr., {Harvey}, J.~W., \emph{{Solar Doppler shifts: Sources of
  continuous spectra}}, in D.~O. {Gough} (ed.) NATO ASIC Proc. 169: Seismology
  of the Sun and the Distant Stars, 105--116 (1986)

\bibitem[{{Pablo} et~al.(2016)}]{2016arXiv160800282P}
{Pablo}, H., et~al., \emph{{The BRITE Constellation nanosatellite mission:
  Testing, commissioning and operations}}, \emph{\pasp}  (2016),
  \eprint{1608.00282}

\bibitem[{{Pigulski} et~al.(2016)}]{2016A+A...588A..55P}
{Pigulski}, A., et~al., \emph{{Massive pulsating stars observed by
  BRITE-Constellation. I. The triple system {$\beta$} Centauri (Agena)}},
  \emph{\aap} \textbf{588}, A55 (2016), \eprint{1602.02806}

\end{thebibliography}

\end{document}